\documentstyle[12pt]{article}

\newcommand{\secn}[1]{Section~\ref{#1}}

\newcommand{\eq}[1]{Eq.~(\ref{#1})}

\def\beq{\begin{equation}}
\def\eeq{\end{equation}}
\def\beqa{\begin{eqnarray}}
\def\eeqa{\end{eqnarray}}
\def\reali{{\hbox{l\kern-.5mm R}}}
\def\compl{{\hbox{l\kern-1.9mm C}}}
\def\interi{{\mathchoice
 {\hbox{Z\kern-1.5mm Z}}
 {\hbox{Z\kern-1.5mm Z}}
 {\hbox{{Z\kern-1.2mm Z}}}
 {\hbox{{Z\kern-1.2mm Z}}}  }}
\newcommand{\sect}[1]{\setcounter{equation}{0}\section{#1}}

\newcommand{\EQ}{\begin{equation}}
\newcommand{\EN}{\end{equation}}
\newcommand{\bea}{\begin{eqnarray}}
\newcommand{\ena}{\end{eqnarray}}

\renewcommand{\a}{\alpha}
\renewcommand{\b}{\beta}

\newcommand{\NP}[1]{Nucl.\ Phys.\ {\bf #1}}
\newcommand{\PL}[1]{Phys.\ Lett.\ {\bf #1}}

\renewcommand{\thefootnote}{\fnsymbol{footnote}}

\begin{document}
\begin{titlepage}
\rightline{DFTT 8/97}
\rightline{\hfill February 1997}

\vskip 1.2cm

\centerline{\Large \bf Scattering of Closed Strings from Many
D-branes}

\vskip 1.2cm

\centerline{\bf Marialuisa Frau, Igor Pesando and Stefano
Sciuto\footnote{e-mail:
sciuto@to.infn.it}}
\centerline{\sl Dipartimento di Fisica Teorica, Universit\`a di
Torino}
\centerline{\sl Via P.Giuria 1, I-10125 Torino, Italy}
\centerline{\sl and I.N.F.N., Sezione di Torino}

\vskip .2cm

\centerline{\bf Alberto Lerda\footnote{II Facolt\`a di Scienze
M.F.N., Universit\`a di Torino
(sede di Alessandria), Italy}}
\centerline{\sl Dipartimento di Scienze e Tecnologie Avanzate and}
\centerline{\sl Dipartimento di Fisica Teorica, Universit\`a di
Torino}
\centerline{\sl Via P.Giuria 1, I-10125 Torino, Italy}
\centerline{\sl and I.N.F.N., Sezione di Torino}

\vskip .2cm

\centerline{\bf Rodolfo Russo}
\centerline{\sl Dipartimento di Fisica, Politecnico di Torino}
\centerline{\sl Corso Duca degli Abruzzi 24, I-10129 Torino, Italy}
\centerline{\sl and I.N.F.N., Sezione di Torino}
\vskip 1cm

\begin{abstract}

We develop an operator formalism
to compute scattering amplitudes of arbitrary bosonic
string states in the background of many D-branes. Specifically, we
construct a suitable boundary state which we use
to saturate the multi-Reggeon vertex in order to obtain the generator  
of
multi-membrane scattering amplitudes.
We explicitly show that the amplitudes with $h$ parallel D-branes are  
similar to amplitudes with $h-1$ open string loops.

\end{abstract}

\end{titlepage}

\newpage
\renewcommand{\thefootnote}{\arabic{footnote}}
\setcounter{footnote}{0}
\setcounter{page}{1}

\sect{Introduction}
\label{intro}

Dirichlet membranes, or simply D-branes, have acquired an increasing
importance in the study of non-perturbative effects in string
theory. As shown in a remarkable paper by
Polchinski \cite{polc}, they provide an exact and simple
conformal field theory description of the extended
solitons carrying one unit of Ramond-Ramond charge in type II
string theories. Since there exist duality transformations  
\cite{hull}
which relate pairs of apparently different string theories
by exchanging the fundamental strings of a model with
the solitons of its dual, it is of some importance to study
the perturbative behavior of a string theory in the background
of D-branes. In fact, from such a perturbative knowledge
one can extract, in principle,
some non-perturbative information on the dual theory.

In this paper we develop an operator formalism that allows
to compute scattering amplitudes of arbitrary string states
in the presence of many D-branes.
Starting from the results already achieved in the old days
of string theory \cite{Ademetal}, we first construct a state
which inserts a boundary on the world-sheet
and enforces Dirichlet or Neumann boundary
conditions on the string coordinates. Then, we take
the multi-Reggeon vertex of the operator formalism
\cite{VNg,phi}, which describes the emission
of many closed strings from a sphere, and saturate it with an
appropriate number of boundary states to compute arbitrary
multi-membrane scattering amplitudes.
Our boundary state is different from those that have
appeared in the literature \cite{Ade2,bounstate,Green,newbounstate}  
since
it contains not only the identity operator that identifies the left
and right sectors of the closed string but also an open string
propagator. It is precisely because of this feature
that our boundary states can be directly used to saturate
the multi-Reggeon vertex of the closed strings to
obtain the multi-membrane amplitudes.

We would like to stress that
the structure of an amplitude with $h$ D-branes
is formally similar to an open string amplitude at $h-1$ loops.
While this analogy is quite obvious from a geometrical point of view,
it is not so straightforward to describe explicitly the interactions  
of
closed strings on surfaces with many boundaries. However,
the operator formalism allows to obtain explicit results
in a closed form.
Even if these calculations are relevant and
interesting in the case of superstrings, where duality
is realized, in this paper we concentrate
on the bosonic string in $D=26$ where it is easier
to understand and display the general structure of the formalism.

To illustrate our procedure, we begin by reviewing in \secn{par1}
the operator formalism for the interaction among open and closed
strings, and extend the old results to the case of Dirichlet
boundary conditions.
In \secn{scucitura}, we factorize the amplitude for the emission of
closed strings from an open string to obtain the boundary operator,  
and,
by sewing two of these, we reproduce Polchinski's
result for the interaction of two parallel D-branes \cite{lpolc}.
Finally in \secn{multibrane} we reformulate the boundary operator as  
a
boundary state, and saturate with it the multi-Reggeon vertex to  
compute
the generator of the multi-brane amplitudes.

\sect{Mixed closed and open string amplitudes}
\label{par1}

The interactions among closed and open strings were extensively
studied already in the early days of string theory
\cite{old,Ademetal}. In particular, in Ref.~\cite{Ademetal}
Ademollo et al. constructed
vertex operators for the emission of a closed string out of an open
string, and computed the scattering amplitudes among $M$ closed
and $N$ open strings at tree level. The topology of the
string world-sheet corresponding to these amplitudes is
that of a disk emitting $N$ open strings
from its boundary and $M$ closed strings from its interior.
As customary in those days, only Neumann boundary conditions
were imposed on the disk, and no target-space compactification was
considered. However, we find useful to recall here
the results of Ref.~\cite{Ademetal} because the introduction of
Dirichlet boundary conditions and of compact directions
is very simple in that approach.

By a conformal transformation a disk can be
mapped onto the upper half
complex $z$-plane ($z={\rm e}^{\tau+i\sigma}$ with $\tau$ and
$\sigma$
being the timelike and spatial coordinates of the world sheet) and
its boundary onto the real axis ($x=\pm{\rm e}^{\tau}$). Then, the
emission from the disk boundary of an open string state with momentum
$p$ and internal
quantum numbers $\alpha$, is described by a vertex operator ${\cal
V}_{\alpha}(x;p)$, while the emission from the
disk interior of a closed string state with momentum
$k$, and left and right quantum numbers $\beta_L$ and $\beta_R$, is
described by a vertex operator
${\cal W}_{\beta_L,\beta_R}(z,{\bar z};k)$.
The presence of a boundary on the world sheet imposes
a relation between the left and right
parts of ${\cal W}_{\beta_L,\beta_R}$ which are not
independent of each other. In fact, if one splits the (Neumann) open
string coordinates
into left and right components
\beq \label{split}
X^{\mu}_{N}(\tau,\sigma)={1\over 2}\left[X^{\mu}(z)
+X^{\mu}(\bar z)\right]~~~,
\eeq
with
\beq \label{splitb}
X^{\mu}(z) = q^{\mu} -{\rm i}\, (2\a')\, a_0^{\mu}\ln z +
{\rm i}\,\sqrt{2\a'}\,
\sum_{n\not =0} {a_n^{\mu}\over {\sqrt {|n|}}}z^{-n} ~~~,
\eeq
it is possible to write \cite{Ademetal}
\beq \label{Wertex}
{\cal W}_{\beta_L,\beta_R}(z,\bar z; k) = {\cal V}_{\b_L}(z;k_L)
{}~{\cal V}_{\b_R}(\bar z; k_R)~~~,
\eeq
where $k_L=k_R=k/2$. We would like to stress that the vertex operator
${\cal W}$ depends on a single set
of oscillators ({\it i.e.} those of the open string), and that each
factor in \eq{Wertex} is
separately normal ordered.
This is to be contrasted with the
vertex operators describing the emission of a closed string
out of a closed string. In fact, in this case there are
two distinct sets of oscillators for the left and right sectors which
only share the zero-mode.

Using the operators ${\cal V}$ and ${\cal W}$,
the tree-level scattering
amplitude among $N$ open and $M$ closed strings
is given by
\bea \label{Ademampl}
A(N, M) & = &{\cal C}_0\,{\cal N}^N \,{\widehat{\cal N}}^M
\int {1\over dV_{abc}}~
\prod_{i=1}^N\left[dx_i\;\theta(x_{i+1}- x_i)
\right]~\prod_{j=1}^M d^2z_j \\ \nonumber
& & \langle 0 |~{\rm T}\left(\prod_{i=1}^N {\cal V}_{\a_i}
(x_i;p_i) \,\prod_{j=1}^M {\cal W}_{\b_{jL},\b_{jR}}
(z_j, \bar z_j; k_j) \right)
| 0\rangle~~~,
\ena
where $dV_{abc}$ is the volume of the projective group $SL(2,{\bf  
R})$,
T denotes the time (radial) ordering prescription, and
${\cal C}_0$, ${\cal N}$ and ${\widehat{\cal N}}$ are respectively
the normalizations of the disk, of the open and of the closed
vertex operators respectively (see {\it e.g.} Ref.~\cite{1loop}).
In \eq{Ademampl} the variables $x_i$'s are integrated on the real
axis while the complex variables $z_j$'s are integrated
on the upper half plane.
Because of \eq{Wertex} it is clear that $A(N,M)$
is formally similar to a pure open string amplitude with
$N+2M$ external states provided suitable identifications
of momenta are made. This has been recently re-proposed
in \cite{Myers}.

It is interesting to note that amplitude (\ref{Ademampl}) is ill
defined if $N=0$.
In fact, as we will see, $A(N,M)$
can be written as a pure closed string diagram with $M+1$ legs one of
which sewn to a disk with $N$ external open strings.
The propagator sewing the closed string amplitude to the
disk must carry a momentum $k$ equal to the sum of the
$N$ open string momenta $p_i$'s; if $N=0$, this sum
vanishes and \eq{Ademampl} becomes ill defined, since the
closed string propagator has a pole when $k=0$.

The situation is different if some target space
directions (labeled by an index $I$) are compactified on a
circle of radius $R$. In this case,
the left and right parts of the closed string
momenta contain a Kaluza-Klein term proportional to $1/R$
and a winding term proportional to $R$:
\beq \label{moment}
{k_L^I} = {1\over 2}\left({n^I\over R}
+ {w^IR\over \a'}\right)~~~,~~~
{k_R^I}  =  {1\over 2}\left({n^I\over R}
- {w^IR\over \a'}\right)~~~.
\eeq
In the compactified theory
\eq{Ademampl} is well defined even if $N=0$. In fact,
it is still true that the amplitude (\ref{Ademampl}) can be
written as a pure closed string diagram sewn to a disk, but now
momentum conservation \footnote{For $N=0$  it is
enforced by a factor $\langle 0|~\exp\Big[{\rm i}\,
q_I\sum\limits_{j=1}^M\left(k^I_{jL}+k^I_{jR}\right)\Big]
| 0\rangle=2\pi R~\delta_{\sum\limits_j n^I_j\,,\,0}$
for any compactified direction.} constrains
only the
Kaluza-Klein part of the sewing
propagator, leaving its winding number arbitrary. Thus,
the singularity is avoided.

Notice that while in a pure closed string amplitude both
the Kaluza-Klein and the
winding numbers are separately conserved,
when the closed strings interact with a disk, only the
conservation of the
Kaluza-Klein part of the momentum seems to hold.
However, contrarily to what happens to the external open strings,
the world-sheet boundary of a virtual open string
can wind along the compact directions, and since
\beq \label{diskwind}
\oint d\sigma^\a~{\partial_\a X^I_N\over 2\pi \a'}  =
\sum_{j=1}^M {w^I_j R\over \a'}~~~,
\eeq
where the contour integral is over the disk boundary,
the conservation of the winding number
can be recovered if one also considers the boundary of the
world sheet.
\eq{diskwind} suggests that
it is necessary to add to the string action of the compactified
Neumann theory the following topological term \cite{Green}
\beq \label{topwind}
{\rm i} \,Y_I
\oint d\sigma^\a~{\partial_\a X^I_N\over 2\pi \a'}~~~,
\eeq
where the parameters $Y_I$, conjugate to the winding number,
are constant $U(1)$ gauge potentials that couple to the
boundary of the disk. Obviously, the term (\ref{topwind})
does not change the equations of motion of the string
coordinates $X^I$ and their mode expansion,
but it couples to the winding number of the external closed
strings so that the amplitude
(\ref{Ademampl}) must be multiplied by the factor
$\exp\Big({\rm i}\,Y_I\sum\limits_{j=1}^M w^I_j \,R/\a'\Big)$.
Such a factor can be automatically generated by shifting
the left and right parts of the open string compact coordinates
$X_N^I$ according to
\beq \label{shift}
X^I(z) \rightarrow X^I(z) + Y^I ~~~~,~~~~
X^I(\bar z)  \rightarrow X^I(\bar z) -Y^I ~~~.
\eeq

Using this approach, it is now rather simple to introduce
Dirichlet boundary conditions along compact directions and study the
scattering amplitudes of closed strings with a D-brane. To do this,
let us first recall that the mode expansion of a compact string
coordinate $X_D^I$ with Dirichlet boundary conditions
$X^I_D(\tau,\sigma=0)=X^I_D(\tau,\sigma=\pi)=Y^I ~{\rm mod }~2 \pi
R$, is
\beq \label{XD}
X_D^I(\tau,\sigma)= Y^I+2w^IR\,\sigma+{\rm i}\, \sqrt{2\a'}
\sum_{n\not =0}{a_n^\mu\over {\sqrt n}}(z^n-{\bar z}^{-n})~~~.
\eeq
As in the Neumann case, also in the Dirichlet case one can divide
the string coordinates into left and right components, and
then follow the same steps as before \cite{Green}.
In fact, using the expansion
(\ref{splitb}), we have
\beq \label{X'}
X_D^I(\tau,\sigma)\equiv {1\over 2}\Big[X^I(z) -X^I(\bar z)\Big]~~~.
\eeq
Notice that the eigenvalues of $a^I_0$ in a compactified Neumann
coordinate $X^I_N$ are $n^I/R$, while in a Dirichlet coordinate
$X_D^I$ they are $w^IR/\a'$ (with $n^I,w^I \in {\bf Z}$).
Thus, the transformation $R \rightarrow \a'/R$, $X^I({\bar z})
\rightarrow -X^I({\bar z})$ and $n^I \rightarrow w^I$
changes Neumann into Dirichlet boundary conditions
in the open strings. Under this duality transformation
the parameters $Y^I$ of the Neumann theory become
the coordinates of the D-brane
in the Dirichlet theory, while the closed string sector remains
unchanged.

We are now in the position of writing the amplitude of closed
strings interacting with a D-brane of dimension $p$.
For notational simplicity we suppose that all 26 coordinates
are compactified with scale $R$ and that the first
$p+1$ have Neumann boundary conditions, while the
remaining $25-p$ have Dirichlet boundary conditions.
Then, the scattering amplitude of $M$ closed string states
with a D$p$-brane is given by
\bea \label{Ddisk}
A(M) &=& V\,{\cal C}_0\,{\widehat{\cal N}}^M~{\rm e}^{{\rm i}\,Y  
\cdot
\sum\limits_{j=1}^M\big({k_{jL} -Sk_{jR} \big)}}
\int{1\over dV_{abc}}~\prod_{j=1}^M \,d^2 z_j
\\
&&\times \langle 0 |~{\rm T}
\left(\prod_{j=1}^M{\cal V}_{\beta_j}(z_j;k_{jL})\,
{\cal V}_{S\beta_{jR}}({\bar z}_j;Sk_{jR}) \right) |0\rangle
{}~~~,\nonumber
\ena
where $S$ is a diagonal matrix with eigenvalues
$+1$ ($-1$) for the Neumann (Dirichlet) directions, and
${\cal V}_{S\beta_R}({\bar z};Sk_{R})$ stands for the
antiholomorphic part of the vertex (\ref{Wertex}) in which
$X({\bar z})$ has been replaced by $SX({\bar z})$.
Note that in \eq{Ddisk}, differently from \eq{Ademampl},
the vacuum $|0\rangle$ has been normalized to one.
This explains the overall appearance of the volume $V$, which,
according to footnote 1, is given in this case by
\beq \label{volume}
V =(2\pi R)^{p+1}~
\left(\frac{2\pi\a'}{R}\right)^{25-p}~~~.
\eeq

The specific expression of the vertices ${\cal V}$ in \eq{Ddisk}
depends on the emitted states,
but it always contains a factor proportional to
$:\exp\Big({{\rm i}\,k_{jL}\cdot X(z_j)}\Big):$
$:\exp\Big({{\rm i}\,Sk_{jR}\cdot X(\bar z_j)}\Big):\,$. Therefore,
the expectation value over
the zero-modes in \eq{Ddisk} yields
the momentum conservation
\beq \label{momcons}
\sum_{j=1}^M\left(k_{jL} + Sk_{jR} \right) = 0
\eeq
{\it i.e.} $\sum\limits_j n^I_j=0$
along the Neumann directions, and $\sum\limits_jw^J_j=0$
along the Dirichlet directions.
We conclude by mentioning that using this formalism
in the decompactification limit $R\to\infty$ one can easily reproduce
the results of Refs.~\cite{Kleb,Myers} for the
amplitudes of massless states from a D-brane.

\sect{Factorization and boundary operator}
\label{scucitura}

We now show that the amplitude $A(M)$ of \eq{Ddisk} can be
factorized as a $M+1$ closed string diagram
in which one leg is saturated with a boundary operator $B$
that encodes the presence of the D-brane. To do this, following
Ref.~\cite{Ademetal}, we first exploit the $SL(2,{\bf R})$ invariance
to fix $z_1 ={\rm i}$, and then make the change of variables
\beq \label{z'}
z \rightarrow z'=-{z-{\rm i} \over z+{\rm i}}~~~,
\eeq
so that the upper half $z$-plane is mapped into the circle of unit
radius.
After this transformation, the variables $z'$ and $(\bar z)'$
are no longer complex conjugate of each other since $(\bar
z)'=1/\overline{z'}$.
Therefore, radial ordering forces to split the vertices
${\cal W}$
into their constituent factors, and put all the
holomorphic
parts on the right and all the antiholomorphic ones on the left.
Then, \eq{Ddisk} becomes
\bea \label{interm1}
A(M) &=& \frac{V\,{\cal C}_0\,{\widehat{\cal N}}^M}{2\pi}
{}~{\rm e}^{{\rm i}\,Y \cdot
\sum\limits_{j=1}^M\big({k_{jL} -Sk_{jR} \big)}}
\int \prod_{j=2}^M \,{d^2 z_j'\over \overline{z_j'}^{\,2}}~~
\langle Sk_{1R};S\b_{1R}|
\\
&&\times\,
{\rm T}\Bigg(\prod_{j=2}^M{\cal
V}_{S\b_{jR}}\!\left(1/\overline{z_j'};Sk_{jR}\right)\Bigg)\,{\rm
T}\Bigg(\prod_{j=2}^M{\cal V}_{\b_{jL}}
(z_j'; k_{jL})\Bigg)
|k_{1L};\b_{1L} \rangle ~~~, \nonumber
\ena
where the states $|k_{1L};\b_{1L} \rangle$ and $\langle
Sk_{1R};S\b_{1R}|$
correspond, respectively, to the vertices
${\cal V}_{\b_{1L}}(z'_1; k_{1L})$ and ${\cal
V}_{S\b_{jR}}\!\left(1/\overline{z_1'};Sk_{1R}\right)$
in the limit $z_1'\to 0$. Notice that the overall factor of $1/2\pi$
in \eq{interm1} is what remains of $1/dV_{abc}$
after fixing $z_1={\rm i}$, {\it  i.e.}
the inverse volume of the translations.
To simplify the notation, from now on
we will suppress the primes on the $z$-variables.

Then, using the relation
${\cal V}_{\beta_j}(z_j)=z_M^{L_0-1}
{\cal V}_{\beta_j}(z_j /z_M)z_M^{-L_0}$, and
inserting a complete set of open string
states $|q;\lambda\rangle$ twice,
\eq{interm1} becomes
\bea
A(M) &=&
\frac{V\,{\cal C}_0\,{\widehat{\cal N}}^M}{2\pi}
{}~{\rm e}^{{\rm i}\,Y \cdot
\sum\limits_{j=1}^M\big({k_{jL} -S {k_{jR}} \big)}}~
\int\prod_{j=2}^M \,{d^2z_j\over {{\bar z}_j^2}}
\left({{\bar z}_M \over z_M}\right)^{M-2}\,
{1 \over z_M^2}\nonumber\\
&&\times
\sum_{\{q;\lambda\},\{q';\lambda'\}}
{}~\langle Sk_{1R};S\b_{1R}|
\,{\rm T} \Bigg(\prod_{j=2}^M{\cal V}_{S\b_{jR}}
\left({\bar z}_M/{{\bar z}_j};Sk_{jR}\right)\Bigg)
|q;\lambda\rangle \label{interm2}\\
&&\times~\langle -q;\lambda|~|z_M|^{2L_0}\,
|q';\lambda'\rangle
\,\langle -q';\lambda'|\,
{\rm T}\Bigg(\prod_{j=2}^M{\cal V}_{\b_{jL}}\left( {z_j/z_M};
k_{jL}\right)\Bigg)| k_{1L};\b_{1L} \rangle ~~~.
\nonumber
\ena
Using the invariance of the second line of \eq{interm2}
under the transformation $X(\bar z)\rightarrow SX(\bar z)$,
that is $(Sk_{jR},S\b_{jR},q, \lambda)$ $\rightarrow$
$(k_{jR},\b_{jR},Sq,S\lambda)$, the transposition property
$\left[{\cal V}_{{\b}_R}\left({1/\bar z}\right)
\right]^{\rm T}
={\bar z}^2 {\cal V}_{\b_R}({\bar z})$, and
the conservation law
$q'=-\sum\limits_{j}k_{jL} =
S\sum\limits_{j}k_{jR}$,
we can rewrite the amplitude $A(M)$ as follows
\beq
A(M) = \sum_{\{q;\lambda\},\{q';\lambda'\}}
 \langle -q';\lambda'|~\langle Sq;S\lambda|\,W\,\times\,
\langle -q;\lambda|\,B\,|q';\lambda'\rangle
\label{factor}
\eeq
where
\bea
W &=& \Phi\,\widehat{{\cal C}_0}\,{\widehat{\cal N}}^{M+1}
\int\prod_{j=2}^{M-1} d^2\xi_j
{}~{\rm T} \Bigg(\prod_{j=2}^M{\cal V}_{\b_{jR}}
\left({\bar \xi}_j;k_{jR}\right)\Bigg)
|k_{1R};\b_{1R}\rangle
\nonumber \\
&&\times {\rm T} \Bigg(\prod_{j=2}^M{\cal V}_{\b_{jL}}
\left( \xi_j;k_{jL}\right)\Bigg)
| k_{1L};\b_{1L} \rangle ~~~,
\label{wclo}
\ena
with $\xi_j=z_j/z_M$, and
\beq
B= \frac{V\,{\cal C}_0}{\Phi\,\widehat{{\cal C}_0}\,{\widehat{\cal  
N}}}
{}~{\rm e}^{-2{\rm i}Y\cdot q'}\int_{|z_M|\leq 1} \frac{d^2z_M}{2\pi}
{}~ |z_M|^{2L_0-4}~~~.
\label{B1}
\eeq
For reasons that will be clear in a moment, we have introduced the  
coefficients
$\widehat{{\cal C}_0}$ and $\Phi$ which are, respectively,
the normalization of the sphere (see {\it e.g.} \cite{1loop})
and the self-dual ``volume'' factor (see {\it e.g.} \cite{Vene})  
which
normalize all closed string scattering amplitudes at tree level.

Eqs. (\ref{wclo}) and (\ref{B1}) can be given a simple  
interpretation.
In fact, we can use two independent sets of oscillators
(say $a_n$ and ${\tilde a}_n$) for the holomorphic and the
antiholomorphic parts of $W$, and relate $Sq$ and $q'$ to the
right and left momenta, $q_R$ and $q_L$,
of a closed string, and $S\lambda$ and $\lambda'$
to its right and left quantum numbers $\lambda_R$ and
$\lambda_L$, according to
\beq
Sq=q_R~~~,~~~q'=-q_L~~~,
{}~~~S\lambda=\lambda_R~~~,~~~\lambda'=\lambda_L~~~.
\label{qlqr}
\eeq
Then, the first factor in \eq{factor}, namely
\beq
{}_{a}\langle q_L;\lambda_L|~{}_{\tilde a}\langle q_R;\lambda_R|
\,W^{a,{\tilde a}}
\label{closampl}
\eeq
can be read as the correctly normalized amplitude among $M+1$ closed
strings in which the first $M$ ones are on shell, and the last
one is on an arbitrary excited state. In this amplitude,
the $SL(2,{\bf C})$ invariance has been fixed by the
conditions $\xi_1=0$, $\xi_M=1$ and $\xi_{M+1}=\infty$; moreover
as is clear from \eq{factor}, the last excited string is sewn to
the operator $B$ which encodes the presence of the boundary.

Since the states $|q;\lambda\rangle$ are eigenstates of
$L_0$, $\lambda=\lambda'$ and
$q=q'$. Therefore, \eq{qlqr} implies that the closed
string states exchanged between $W$ and $B$
satisfy the conditions $\lambda_L=S\lambda_R$
and $q_L=-Sq_R$,
that is $n=0$ ($w=0$)
along the Neumann (Dirichlet) directions. Thus, the explicit form of
the boundary operator is
\footnote{From closed string unitarity, one can derive the relation
$\widehat{{\cal C}_0}\widehat{\cal N}^2\a'=4\pi$; thus the prefactor  
of
\eq{B1} can be written as in \eq{BounOp1}.}
\beq \label{BounOp1}
B = {V\,{\cal C}_0 {\widehat{\cal N}}\a' \over 4\pi\Phi }~
\frac{1}{2L_0-2}~\left.
\prod_{I=0}^p \left({\rm e}^{-{\rm i}\,Y_I w^IR/\a'}\,\delta_{n^I,0}
\right)
\prod_{J=p+1}^{25} \left({\rm e}^{-{\rm i}\,Y_J  
n^J/R}\,\delta_{w^J,0}\right)
\right|_{\lambda_L=S\lambda_R}\!\!.
\eeq
It is important to realize that the operator $B$ contains
the information
about the presence of a D$p$-brane with coordinates $Y_J$,
as well as
a closed string propagator with the level matching condition  
enforced.
This fact is also showed by the normalization of \eq{BounOp1}:
in fact, $V {\cal C}_0$ is the factor common to all
amplitudes having
a disk with no holes as world-sheet, $\widehat{\cal N}$
signals the emission of a closed string state, and
$\a'/\Phi$ gives the right dimension to the closed string
propagator attached to the disk.
As we will see in the next section,
it is the presence of the closed string propagator in $B$
that makes it easy to write
scattering amplitudes with many D-branes in the operator
formalism.

The geometrical meaning of the operator $B$ is quite clear.
For example, if we
sew $B$ to a graph with one closed and
$N$ open strings as in \eq{Ademampl}, we obtain a $N$-point
open string amplitude at one loop. Thus, as expected, $B$ inserts
a boundary on the world sheet, transforming the disk into an annulus
\cite{Ade2}.
The latter turns out to be described in the ``crossed channel'',
where the role of $\tau$ and $\sigma$ is exchanged and the modulus  
$t$
measures essentially the distance between the two boundaries. Making
the modular transformation $t\to 2\pi/t$, one can obtain the one-loop
amplitude in the usual configuration as it results when one computes  
the trace
of open string vertex operators (see chapter VIII of  
Ref.~\cite{GSW}).
Furthermore, by carefully comparing the normalization coefficients
in the two descriptions,
one finds that
\beq \label{op-cl}
\widehat{\cal N} = {\cal N}
\sqrt{\frac{8\pi\Phi\,{\cal C}_1}{V\,{\cal C}_0}}~~~,
\eeq
where ${\cal C}_1=(2\pi)^{-26}(2\a')^{-13}$ is the normalization of  
the annulus
\cite{1loop}.

As a check, let us now consider two boundary
operators $B_1$ and $B_2$ and sew
them together to calculate
the interaction between two parallel
D$p$-branes that exchange closed string states.
Since both operators include a propagator,
their sewing must be done using the
inverse propagator, namely
\beq \label{Btens}
{\cal A}\big|_{R} = {\rm Tr'}\left(B_1^\dagger ~
\left( \frac{2\Phi}{\a'}
(L_0+{\tilde L}_0-2)\right)~B_2\right)~~~.
\eeq
where ${\rm Tr'}$ means trace over the physical states;
this amounts to change the space-time dimension $D$ into $D-2$
in the trace over the non-zero modes.
Thus, with this prescription, one obtains
\bea \label{Tint1}
{\cal A}\big|_{R} &=&
\frac{V\,{\cal C}_1}{2\pi}
{}~\int_0^\infty
dt~{\rm e}^t~\prod_{n=1}^\infty\left(1-{\rm e}^{-nt}\right)^{-24}
\\ \nonumber
&&\times
\prod_{I=0}^p\sum_{w^I}
\exp{\left[-t\a'\left({w^IR\over 2\a'}\right)^2+
2{\rm i} (Y_1-Y_2)_I\left(
{w^IR\over 2\a'}\right)\right]}
\\ \nonumber
&&\times
\prod_{J=p+1}^{25}\sum_{n^J}\exp{\left[-t\a'\left({n^J\over  
2R}\right)^2+ 2{\rm
i}
(Y_1-Y_2)_J \left({n^J\over 2R} \right) \right]}~~~.
\ena
It is now easy to take the decompactification limit  
$R\rightarrow\infty$
of \eq{Tint1}:
in this case the sum over $w$ simply picks up the value
$w\!=\!0$, while the sum over $n$ becomes a gaussian integral, so  
that
using \eq{op-cl}, one gets
\beq \label{Tint2}
{\cal A} =\lim_{R\to \infty} \frac{V\,{\cal C}_1}{2\pi}
{}~\int_0^\infty dt~
\left(\frac{4\pi R^2}{\a'\,t}\right)^{(25-p)/ 2}
\left(f_1({\rm e}^{-t/2})\right)^{-24} \exp{\left[-{
\Delta Y^2\over \a' t}\right]}~~~,
\eeq
where $f_1(q)\equiv q^{1/12}
\prod_{n=1}^\infty\left(1-q^{2n}\right)$ ,
and $\Delta Y^2\equiv \sum\limits_{J=p+1}^{25}  
(Y_1-Y_2)_J\,(Y_1-Y_2)^J$
is the square of the distance between the two D-branes.

By making the modular transformation $t\to 2\pi/t$,
this same amplitude can be reinterpreted as the
one-loop free energy of an open string whose ends
are fixed on two parallel D-branes.
In fact, using  the relation $\left(f_1({\rm e}^{-\pi/t})
\right)^{-24}\!\! =t^{-12}\left(f_1({\rm e}^{-\pi t})\right)^{-24}$
and the explicit expression of ${\cal C}_1$, \eq{Tint2} becomes
\beq \label{Pol}
{\cal A} =V_{p+1} \int_0^\infty {dt\over t}~(8\pi^2\a't)^{-{(p+1)/  
2}}
\left(f_1({\rm e}^{-\pi t})\right)^{-24} \exp{\left[-{
\Delta Y^2\over 2\pi\a'}t\right]}~~~,
\eeq
where $V_{p+1}=(2\pi R)^{p+1}$ is the world volume of the  
D$p$-branes. \eq{Pol}
agrees completely, including the overall normalization,
with the result of Ref.~\cite{lpolc}. By analyzing the poles of
${\cal A}$ due to graviton and dilaton exchanges, one can obtain
the tension of the D-brane which is the normalization of its  
world-volume
action.

The same calculation can be done for two non-parallel D-branes,
even of
different dimensionality; the main change in
Eqs. (\ref{Tint1})-(\ref{Tint2}) is
that, in the directions parallel to
a D-brane and ortogonal to the other, $f_1(q)$ is replaced by
$f_2(q)= q^{1/12}\prod_{n=1}^\infty (1+q^{2n})$ and the zero modes of  
the
exchanged
closed string are forced to be zero.

\sect{Scattering amplitudes with many D-branes}
\label{multibrane}

The purpose of this section is to compute the interaction
among $M$ closed strings in the presence of $h$ parallel D$p$-branes.
{}From the geometrical point of view, this process is
associated to a world-sheet $\Sigma_{M;h}$ with
$M$ punctures and $h$ boundaries on which $25-p$ string coordinates
have Dirichlet boundary conditions.
The surface $\Sigma_{M;h}$ can be obtained
from a sphere with $M+h$ punctures in which
$h$ holes are cut around $h$ punctures.
This operation has an explicit realization in the
operator formalism used in Refs.~\cite{VNg,phi}
for the calculation of multiloop string amplitudes.
In this formalism the fundamental object
is the multi-Reggeon vertex that
generates all scattering amplitudes among arbitrary
string states, and encodes all geometric information
about the corresponding world-sheet. In particular,
to a sphere with $M+h$
punctures one associates a tree-level multi-Reggeon
vertex with $M+h$ closed strings, which we denote by ${\bf V}_{M+h}$.
The second fundamental ingredient that is necessary for
our purpose is the boundary state $|B\rangle$
that inserts a hole around a given puncture and enforces
the appropriate boundary conditions on the string coordinates.
Once $|B\rangle$ is given, the scattering amplitude
of $M$ closed strings in the background
of $h$ D-branes can be obtained by saturating ${\bf V}_{M+h}$
with $M$ on-shell closed string states and $h$ boundary states.
The generator of all such amplitudes is then
the vertex operator
\beq
{\bf V}_{M;h} \equiv {\bf V}_{M+h}~
\prod_{\mu=0}^{h-1}|B_\mu\rangle
\label{vmh}
\eeq
whose corresponding world-sheet is $\Sigma_{M;h}$.
Since this surface is conformally equivalent to
a disk with $h-1$ holes and $M$ punctures, one can guess
that ${\bf V}_{M;h}$ will be similar in its structure to
the Reggeon vertex of the open string at $h-1$ loops \cite{VNg}.
We remark that one can construct ${\bf V}_{M;h}$
for any $M$ and $h$, since all legs of ${\bf V}_{M+h}$
are off shell. Futhermore,
as emphasized in Ref.~\cite{phi}, one of the distinctive
features of the operator formalism is that no
knowledge of $\Sigma_{M;h}$
is a priori necessary; in fact all geometrical objects of
the world-sheet are the result of
the sewing precedure and are given explicitly in the Schottky  
representation.

Let us now give some details. The tree-level multi-Reggeon vertex
for the compactified closed string is
\cite{phi}
\bea \label{close-vert}
{\bf V}_{M}&=&\Phi\widehat{{\cal C}_0} \widehat{\cal N}^M
{}~{\bf V}_M^{(gh)}~
\prod_{i=1}^M\left[\sum_{n^i,w^i}\langle n^i,w^i;0| \right]
\int \prod_{i=1}^M \left(\frac{d^2z_i}{|z_{i+1}-z_i|^2}\right)\!
{1\over d{\widehat V}_{abc}}
\nonumber \\
&&\times~
\delta\Big({\sum\limits_{i=1}^M a_0^{i}}\Big)\,
\delta\Big({\sum\limits_{j=1}^M \tilde a_0^{j}}\Big)
{}~\exp\!\!{\left[-\!\!\sum_{i< j=1}^M\sum_{k,l=0}^\infty \!a_k^{i}
D_{kl}(\Gamma
V^{-1}_i V_j)\cdot a_l^{j} \right]}
\nonumber\\
&&\times~
\exp\!\!{\left[-\!\!\sum_{i< j=1}^M\sum_{k,l=0}^\infty\!{\tilde  
a}_k^{i}
D_{kl}(\Gamma \bar V^{-1}_i\bar V_j)\cdot{\tilde a}_l^{j} \right]}~.
\ena
In this expression, $V_i$ is a projective transformation related to
the choice of the local coordinates around the puncture $z_i$
such that $V_i(0)=z_i$, $\Gamma$ is the
transformation $z \rightarrow 1/z$, the matrices $D_{nm}$ are the
infinite dimensional representation of the projective group of
 weight zero, and $|n, w;0\rangle$ is an eigenstate of $a_0$ and
${\tilde a}_0$ with eigenvalues $k_L$ and $k_R$ as in \eq{moment}.
Finally, the variables $z_i$ are integrated over the whole complex
plane and $d{\widehat V}_{abc}$ is the volume of the
projective group $SL(2,{\bf C})$.
For sake of simplicity we do not write the ghost vertex
${\bf V}_{M}^{(gh)}$ which can be found in Ref.~\cite{VN0}.
Note that \eq{close-vert} exhibits a complete left-right  
factorization and
its structure is formally similar to the product
of two independent open string vertices.

Let us now turn to the boundary state $|B\rangle$.
To obtain it, we start from the boundary operator
of the previous section which, however, must be modified
for two reasons. In fact, the operator $B$ of
\eq{BounOp1} geometrically represents
a closed string propagator attached to a disk, and is formally
composed by two parts:
the first one identifies the left and the right sectors
of a closed string, and the second is an open string
propagator that sews them together \footnote{This explains
why the operator $B$ brings only {\it one} real
modulus, even if it contains a closed string propagator.}.
In \secn{scucitura}, the identification was performed
by transforming one of the two bras of \eq{closampl}
into a ket, and then the sewing was realized by taking the trace
on the open string propagator.
Now the situation is slightly different: in fact,
since the state attached to the boundary operator is not
necessarily fixed at $z=0$ or $z=\infty$, it
is represented by a vertex ${\cal W}$,
and not simply by a bra or a ket.
Thus, to identify the two sectors of the closed string
we have first to take the adjoint of, say, the right
part of ${\cal W}$.
This operation contains a twist
since the transformation property of the vertices is
${\cal V}_{\b_R}^{\dagger}({1/\bar z})
=z^2 (-1)^{\ell_R}{\cal V}_{\b_R}(z)$,
where $\ell_R$ is the level of the state. If one
wants that the closed string ends on a membrane (and not on a
crosscap), this twist has to be removed. This can be done
by sewing the two sectors of the closed string by means
of a {\it twisted} propagator.

The second modification of $B$ stems out of the presence of
unphysical states.
In \secn{scucitura} they have been projected out by hand, but
in the background of three or more D-branes this is not possible,
and thus one has to include ghosts and work with BRST
invariant objects.
Since the boundary conditions of the string coordinates
do not influence the local geometry on the world-sheet,
the reparametrization ghosts do not feel the boundary conditions,
and are the same as in the standard Neumann open string theory.
Therefore, the propagator $1/(L_0-1)$ in
\eq{BounOp1} must be replaced by a
BRST invariant twisted open string propagator $T$.
Its explicit
form depends on the local coordinates $V_i$ around the punctures
that are sewn together; for example,
if Lovelace coordinates are used, we have \cite{VN0}
\beq \label{propag}
T= (b_0-b_1) \int_0^1 \frac{dx}{x(1-x)} P(x)
\eeq
where $P(x)$ is the operator that realizes the
transformation $z\rightarrow(xz-x)/(xz-1)$~
\footnote{Note that this transformation contains a twist since its  
determinant
is negative.}, and
$b_0$ and $b_1$ are the antighost zero-modes.
(For the ghost fields, here and in the following we adopt the  
notations of
Ref.~\cite{FMS}).
By attaching this propagator to the operator that
identifies the left and right sectors of the
closed string \cite{Ade2,bounstate,Green,newbounstate} one
gets the BRST invariant state
\beq \label{general}
|B\rangle =\sqrt{V\,{\cal C}_1\a' \over 2\pi\Phi }\int_0^1 {dx\over
x(1-x)} |B(x)\rangle_{X} |B(x)\rangle_{gh}~~,
\eeq
where
\bea \label{X}
|B(x)\rangle_{X} &=& \exp{\left[-\sum_{n,m =0}^\infty a_n^\dagger
D_{nm}(P(x))\cdot S\tilde{a}_m^\dagger\right]}
\\ \nonumber &&\times\;
{\rm e}^{-iY\cdot (a_0-S\tilde a_0)}
\prod_{I=0}^p\left (\sum_{w^I} |0,w^I;0\rangle\right)\!\!
\prod_{J=p+1}^{25}\left (\sum_{n^J}|n^J,0;0\rangle\right)~~,
\ena
and
\bea\label{ghost}
|B(x)\rangle_{gh} &\!\! = &\!\! (b_0-b_1)
\exp\!{\left[\sum_{k,l=-1}^{\infty}
c_k^\dagger E_{kl}(P(x)) \tilde b_l^\dagger
+\sum_{m,n=2}^{\infty} \tilde c_m^\dagger
E_{mn}(P(x)) b_n^\dagger\right] } \\ &\!\! &\!\!\nonumber \times
\exp\!{\left[-\sum_{n=2}^\infty \sum_{r,s=-1}^{1} \tilde c_n^\dagger
E_{nr}(P(x))E_{rs}(\Gamma P(x)) \tilde b_{s}^\dagger\right] }
|q=0\rangle |\tilde q=3\rangle~~~,
\ena
with $E_{nm}$ being the infinite dimensional representation
of the projective group of weight $-1$ \cite{VN0}.
Notice that this structure of the
boundary state $|B\rangle$ is completely general; if different
local coordinates are used, the only thing that changes is the  
explicit form of
the twisted propagator $T$.
The operator $|B(x)\rangle_{X}$ satisfies the equations
\beq
\left( a_n + \sum_{m=0}^\infty D_{nm}(P(x))
S\,{\tilde a}_m^\dagger \right)|B(x)\rangle_{X} = 0
\label{b}
\eeq
with $n>0$, and
\beq
\left(a_0 + S\,{\tilde a}_0\right) |B(x)\rangle_{X} = 0~~~.
\label{bg}
\eeq
Analogous equations hold for $|B(x)\rangle_{gh}$.
We remark that the boundary states considered in the literature
\cite{Ade2,bounstate,Green,newbounstate} satisfy Eqs. (\ref{b})
and (\ref{bg}) with $D_{nm}$ replaced by $\delta_{nm}$.

It is easy to check that by saturating one leg of ${\bf V}_{m+1}$
of \eq{close-vert} with $|B\rangle$, one obtains an
operator
\beq \label{Vb>}
{\bf V}_{M;1}={\bf V}_{M+1}\,|B\rangle ~~,
\eeq
which generates the amplitudes (\ref{Ddisk}) when furtherly
saturated with $M$ physical closed string states.
Notice that the insertion of the boundary soaks
up three real parameters of $SL(2,{\bf C})$ leaving
a residual $SL(2,{\bf R})$ invariance.

We have now all ingredients to compute ${\bf V}_{M;h}$;
it is only a matter of saturating ${\bf V}_{M+h-1;1}$ with
further $h-1$ boundary states, and
of using the oscillator algebra to calculate the
expectation values in \eq{vmh}.
In the ghost sector, this task is simplified by
saturating also the remaining $M$ legs with the
ghost part of the external physical states, which, due
to BRST invariance, can always be written as
$c_1\,{\tilde c}_1|q={\tilde q}=0\rangle$.
As we have
already stressed, the calculation over the orbital
part gives the geometrical objects of the surface
in the Schottky representation, while
the ghost sector simply modifies the measure of
integration. Note that the presence of Dirichlet or Neumann boundary  
conditions
does not alter the result of the traces on the
non-zero modes.
In fact, after the insertion of the first $|B\rangle$
(\eq{Vb>}) the right part of all external states
appears multiplied by the factor $S$, see \eq{Ddisk}.
When further boundaries are introduced, one identifies
the right and the left part of the string field by means of a  
propagator that
contains another factor $S$, see for example \eq{X}. Thus, only  
$S^2=1$ appears
and the calculation of the traces is the same as in the
standard open string
theory (see Ref. \cite{phi} for details).
Evidence of this fact already appeared in \secn{scucitura}:
indeed, the trace in \eq{Tint1} yielded the
usual partition function $f_1(q)$ characterizing
the one-loop amplitudes of the open string for all
24 transverse directions independently of the boundary
conditions. Thus, as long as the matrix $S$ is equal for
all the boundaries, {\it i.e.} the D-branes are parallel,
a sphere with $h$ boundaries yields the same geometrical objects that  
describe
an amplitude of open string with $h-1$ loops.

We now concentrate on the decompactification limit $R\rightarrow
\infty$ and write explicitly ${\bf V}_{M;h}$ in terms of the prime  
form
$E(z_i,z_j)$, the first abelian differential $\omega(z)$ and the  
period
matrix $\tau_{\mu\nu}$ of the surface $\Sigma_{M;h}$.
As we have already seen in the previous section, in this limit
the only possible value for the winding number is $w=0$ and thus the
zero modes $a_0$ and ${\tilde a}_0$ have to be identified,
while the structure of the orbital vacua in \eq{X} becomes
\beq \label{continuo}
\prod_{I=0}^p\left (\sum_{w^I} |0,w^I;0\rangle\right)\!\!
\prod_{J=p+1}^{25}\left (\sum_{n^J}|n^J,0;0\rangle\right)
\rightarrow\prod_{I=0}^{p}|0;0\rangle\!
\prod_{J=p+1}^{25}\!\left(\int  |q^J;0\rangle dq^J\right )~~.
\eeq
In order to write ${\bf V}_{M;h}$ in a more compact form,
we fix $2\a'=1$, neglect the normalization factors
and introduce a new convention for the oscillators
\beq \label{alpha}
\a^{i}_n =
\left\{\matrix{
 \!\!\sqrt{n}\,a^{i}_n \cr \cr
 \sqrt{n}\, S\,\tilde a^{2M-i+1}_n \cr}\right.
{}~~,~~
\a^{i}_0 =
\left\{\matrix{
 \!\!\!\!a^{i}_0~~~~~~~~~~~{\rm for}~i=1,\ldots,M\,, \cr \cr
 Sa^{2M-i+1}_0 ~~~~~~~~{\rm for}~i=M+1,\ldots,2M\,.\cr}\right.
\eeq
Then, after integrating over the loop momenta $q^J$ along the
Dirichlet directions, and setting $Y_0=0$,
the Reggeon vertex ${\bf V}_{M;h}$ for the orbital degrees
of freedom can be written as
\bea
{\bf V}_{M;h}&\!=\!&\label{manybrane}
\prod_{i=1}^M\left(\int dp_i\langle p_i;0|\right)\,
\delta^{p+1}\left(\sum_{i=1}^M \a_0^{i}\right)\int
dm_h~({\rm det\; Im}\,\tau)^{~(p-25)/2}
\\ \nonumber && \times
\exp\left(\sum_{i<j=1}^{2M}\sum_{n,m=0}^\infty
\frac{\a_n^{i}}{n!}
\partial_z^n \partial_y^m
\log E(V_i(z),V_j(y))\Big|_{z=y=0}
\frac{\a_m^{j}}{m!}\right)
\\ \nonumber && \times
\exp\left(\frac{1}{2}\sum_{i=1}^{2M}\sum_{n,m=0}^\infty
\frac{\a_n^{i}}{n!}
\partial_z^n \partial_y^m
\log{E(V_i(z),V_i(y))\over V_i(z)-V_i(y)}
\Bigg|_{z=y=0}
\frac{\a_m^{i}}{m!}\right)
\\ \nonumber &&\times
\prod_{J=p+1}^{25}\exp\!
\left\{{1\over 2}\sum_{\mu,\nu=1}^{h-1}\Bigg[
\Bigg(\sum_{i=1}^{2M}\sum_{n=0}^\infty
\frac{\a_n^{i}}{n!}\partial_z^n \int_{z_0}^{V_i(z)}\!\!
\omega^{\mu}\Bigg)\!\Bigg|_{z=0} \!\!\!\!\!\!\! - 2{\rm  
i}Y^\mu\Bigg]_J
\right.
\\ \nonumber && \times \left.
(2\pi {\rm Im}\tau)^{-1}_{\mu\nu}
\Bigg[\Bigg(\sum_{j=1}^{2M}\sum_{m=0}^\infty
\frac{\a_m^{j}}{m!}\partial_y^m \int_{z_0}^{V_j(y)}\!\!
\omega^{\nu}\Bigg)\!\Bigg|_{y=0}\!\!\!\!\!\!\! -2{\rm i}Y^\nu\Bigg]_J
\right\}~~,
\ena
where the local coordinates $V_i$ of the external states have been  
chosen as
\beq \label{z+zi}
V_{i}(z) =
\left\{\matrix{\!\!\!\!\!\!\!\!
 z+z_i~~~~~~~~~~~{\rm for}~i=1,\ldots,M\,, \cr \cr
 z+\bar z_{2M-i+1} ~~~~~~~~{\rm for}~i=M+1,\ldots,2M\,,\cr}\right.
\eeq
and the measure $dm_h$, that already takes into account the ghost  
contributions
\cite{VNg}, is
\beq \label{mis}\!\!\!
dm_h = {\prod_{i=1}^M d^2 z_i\over dV_{abc}}\prod_{\mu=1}^{h-1}
\left({dk_\mu d^2\eta_\mu (1-k_\mu)^2\over
k_\mu^2(\eta_\mu-\bar\eta_\mu)^2}\right)
{\prod_{\a}}'\left(\prod_{n=1}^\infty (1-k^n_\a)^{-D}
\prod_{n=2}^\infty (1-k^n_\a)^2\right)~.
\eeq
The parameters in \eq{mis} are the moduli of the surface  
$\Sigma_{M;h}$
in the Schottky representation, namely a real multiplier $k_{\mu}$
 and two complex conjugate fixed points $\eta_{\mu}$ and  
$\bar\eta_{\mu}$
for each of the $h-1$ generators of the Schottky group.
The explicit expressions of the  prime form, abelian differentials
and period matrix in terms of the Schottky parameters can be found
in Ref.~\cite{phi}. The vertex ${\bf V}_{M;h}$ depends on  
$2M+3(h-1)-3$
moduli, that is exactly what one expects, because of the close  
analogy
between multi-brane processes and multi-loop open string amplitudes.
Note that for $M=0$ and $h=2$ \eq{manybrane} gives the same result
for the interaction of two D-branes found in \secn{scucitura}.
It is interesting to note that, along the Dirichlet directions,
the bilinear part of \eq{manybrane} with $i \neq j$
can be rewritten in terms of the Dirichlet Green function used in
the functional approach. If $i =j$, where the Green function is  
singular
and the functional approach requires a suitable regularization  
\cite{Gutperle},
our operator formalism already gives a well-defined expression  
because
the closed string emission vertex is separately normal ordered in the
left and the right parts.

We end by commenting that our formalism is suitable for several
extensions and applications; in particular it is possible to take  
into
account the emission of open strings from the boundaries,  consider  
the
case of many non parallel D-branes or boost the boundary states. We  
plan to
extend this formalism to the
superstring and compute the field theory limit of the
multi-membrane scattering amplitudes.
Recently some of these issues have been addressed in  
Ref.~\cite{Iengo}.

\vskip 0.5cm
\noindent
{\bf Acknowledgements}

\noindent
This work is partially supported by the European Commission
TMR programme ERBFMRX-CT96-0045 in which R.R. is associated to Torino
University.

\end{document}